\documentclass[a4paper, 11pt]{article}
\usepackage[english]{babel}
	\addto{\captionsenglish}{}
\usepackage{amsmath,amsthm}
\usepackage{amsfonts}
\usepackage[dvips]{graphics}
\usepackage{graphicx,latexsym,amssymb,amsmath,cancel,color,enumerate,multirow,lscape,bm,rotating}
\usepackage{listings}
\usepackage[svgnames]{xcolor}
\usepackage[pass]{geometry}
\usepackage{pdflscape}
\usepackage{caption}
\usepackage{soul}
\usepackage{bbm}
\usepackage[all]{xy}
\usepackage[square]{natbib}
\usepackage{setspace}
\usepackage[section]{placeins}	
\usepackage{float}	
\usepackage[T1]{fontenc}
\usepackage{lmodern}
\usepackage[nottoc,notlot,notlof]{tocbibind}
\usepackage{hyperref}

\newcommand{\proglang}{\texttt}
\newcommand{\pkg}{\texttt}
\newcommand{\fct}{\texttt}
\newcommand{\code}{\texttt}

\usepackage{graphicx,scalerel,mathtools}

\ifx\theorem\undefined
\newtheorem{theorem}{Theorem}

\begin{document}

\begin{center}

HCmodelSets: AN R PACKAGE FOR SPECIFYING SETS OF WELL-FITTING MODELS IN REGRESSION WITH A LARGE NUMBER OF POTENTIAL EXPLANATORY VARIABLES.\\

\bigskip

By\\

\medskip

H.~H.~Hoeltgebaum\\
{\it Pontifical Catholic University of Rio de Janeiro} and \\
{\it Department of Mathematics, Imperial College London}\\

\medskip

and

\medskip

H.~S.~Battey\\
{\it Department of Mathematics, Imperial College London}\\

\medskip

\end{center}
\begin{center}
SUMMARY \\
\end{center}
In the context of regression with a large number of explanatory variables, Cox and Battey (2017) emphasize that if there are alternative reasonable explanations of the data that are statistically indistinguishable, one should aim to specify as many of these explanations as is feasible. The standard practice, by contrast, is to report a single model effective for prediction. The present paper illustrates the \proglang{R} implementation of the new ideas in the package \pkg{HCmodelSets}, using simple reproducible examples and real data. Results of some simulation experiments are also reported. \vspace{6mm}

\noindent Some key words: confidence sets of models; genomics; regression analysis; sparse effects.

\section{Introduction}

In a recent paper Cox and Battey (2017) outline a procedure for regression analysis when there are more explanatory features than study individuals, a situation that arises particularly in genomics. Their emphasis is on understanding of the true data generating mechanism rather than prediction. The distinction is important. For prediction there may be several models that are essentially equally effective and any choice between them is rather arbitrary. On the other hand, since different well-fitting models typically have different subject-matter implications, it is insufficient, and often misleading, to report an arbitrary one. Even if the immediate goal is prediction, a causal explanation is likely to produce more stable and more generalizable predictions. A key message of Cox and Battey (2017)  is that if there are several models that fit the data essentially equally well, one should aim to specify as many as is feasible. This view is in contraposition to that implicit in the use of the lasso (Tibshirani, 1996) and other variable selection methods, which produce a single model effective for prediction.

The methods of Cox and Battey (2017)  are summarized in \S \ref{sec:methods}. Software implementing these ideas in \proglang{R} has been written by Hoeltgebaum (2018) and is available at \url{https://cran.r-project.org/web/packages/HCmodelSets/} in the package `HCmodelSets'. The software supports most widely used models of dependency including the normal theory linear model, the linear logistic model for binary data (Cox, 1958), and the proportional hazards model fitted by partial likelihood (Cox, 1972; Cox, 1975b). The present article aims to provide a detailed guide to usage based on simple examples. 

\section{Methodology} \label{sec:methods}

Suppose that data are available on $n$ units, for each of which an outcome $y$ is observed along with a vector $x$ of $d$ potential explanatory variables, where $d$ is much larger than $n$. For progress an assumption of sparsity is needed, and the most explicit and interpretable such assumption is that relatively few of the potential explanatory features have a real effect, an assumption central to the formulation of the lasso and similar penalized regression procedures.


Cox and Battey (2017) suggest a different approach whose aim is essentially a confidence set of models. There are three stages to a discussion, and conditionally on the first two, the resulting set of models can be made to have the formal statistical properties associated with confidence sets. 

In the first stage, an initial reduction is made in which a large number of variables are discarded on the basis that they have no explanatory power, or that any explanatory power that they appear to have is explained away by other variables. The assessment is made by fitting a suitable low-dimensional regression model several times to each variable, each time alongside a different set of $k$ companion variables. A variable is retained for further study if it satisfies a particular criterion in at least half of the analyses in which it appears. The sets of variables to be considered together are specified by a partially balanced incomplete block design (Yates, 1936) in which variable indices are arranged in a hypercube of appropriate dimension. This initial dimension is determined by $d$ and a constraint on $k$ to mitigate the effect of dependence between $p$-values, or the associated test statistics, in any single analysis. Ideally $k$ will be between 10 and 15;  see \S 7 of Battey and Cox (2018) for a discussion of this choice. Successive reductions are made using arrangements in successively lower-dimensional hypercubes, where the criterion for retaining variables in each stage is guided by the theoretical discussion of Battey and Cox (2018), the need to keep the number of rows, columns, etc., of successive hypercubes ideally $\leq 15$, and the requirement for a degree of stability over rerandomization of the variable indices in the hypercube. Thus, judgement is required at various stages.

On the resulting set of variables, of which there will be roughly 10-20 by construction, an exploratory analysis is performed, of the kind that is standard in much statistical work. For instance, inspection of interaction plots or probability plots of $t$ statistics. The objective is to detect possible nonlinearities or outliers.

All variables retained through the reduction phase and any squared or interaction terms suggested at the exploratory phase comprise the so called comprehensive model. All low dimensional subsets of the comprehensive model are then tested for their compatibility with the data using a likelihood ratio test, and all models that pass this test are reported. If, among this set, there are models that contain interaction terms without the corresponding main effects, the main effects are added.

For the resulting sets of models to have the formal statistical properties associated with confidence sets, conditional on the first two phases, it is necessary to either split the sample, see Cox (1975a) for a discussion, or to adjust standard tests of model adequacy in account of the alternative hypothesis being selected in the light of the data.

\section{Illustration of usage: a simple reproducible example} \label{sec:illustrations}

\subsection{Some simple data generating processes}\label{sec:dgp}

We illustrate the functions available in `HCmodelSets', and their appropriate usage, using simple examples. These functions include \fct{DGP}, which can be used to reproduce the simulation study of Battey and Cox (2018) and to explore further sensitivities.

\begin{lstlisting}
R > library(HCmodelSets)
R > dgp = DGP(s=5, a=3, sigStrength=1, rho=0.9, n=100, 
       intercept=5, noise=1, var=1, d=1000, DGP.seed = 2018)
\end{lstlisting}
This generates normally distributed responses as $Y_{i}=\mu+x_{i}^{T}\beta + \varepsilon_{i}$  ($i=1,\ldots, n$) where, in the present example \code{n = 100}, $\mu\texttt{ = intercept = 5}$, the $\varepsilon_{i}$ are independently standard normally distributed and the $x_{i}$ are realizations of a \code{d = 1000} dimensional normal random vector of mean zero and covariance matrix $\Sigma$. The vector of regression coefficients $\beta$ is sparse in the sense that only \code{s = 5} entries are non-zero, equal to \code{sigStrength = 1}, and $\Sigma$ is such that a correlation \code{rho = 0.9} is induced between the corresponding entries of $x_{i}$, the so called signal variables, and among \code{a = 3} of the remaining variables. All potential explanatory variables have variance \code{var = 1}.  The results of the subsequent analysis can be reproduced by setting \code{DGP.seed = 2018} as above, but this argument is not needed.

With the appropriate modification to its arguments, the \fct{DGP} function also generates survival times from a proportional hazards model with Weibull baseline hazard. The hazard function for the $i$th individual is
 \[
 h_i(t;\beta)=\kappa \tau(\tau t)^{\kappa-1}\exp\{x_{i}^{T}\beta\},
 \]
where $h(t)=\kappa \tau(\tau t)^{\kappa-1}$ is the Weibull hazard function. 
From this, the density and distribution functions of survival times conditional on $x_{i}$ are obtained as
\begin{eqnarray*}
f_{i}(t;\beta)&=&\kappa \tau (\tau t)^{\kappa -1}\exp\{x_{i}^{T}\beta\}\exp\{-e^{x_{i}^{T}\beta}(\tau t)^{\kappa}\}, \\
F_{i}(t;\beta)&=& \exp\{-e^{x_{i}^{T}\beta}(\tau t)^{\kappa}\}.
\end{eqnarray*}

Thus, given covariates $x_{i}$, uncensored survival times from the above proportional hazards model are generated as $T_{i}=\{-\log U/(\tau^\kappa e^{x_{i}^{T}\beta})\}^{1/\kappa}$, where $U$ is a uniform random variable on $(0,1)$.

In \S \ref{sec:simulations} a minor modification to the previous code is given to generate (potentially censored) survival time data from this model. Simulation results for the procedure fitted to both types of data are also reported in \S \ref{sec:simulations}.

\subsection{Reduction phase}

Based on the previous output, typical usage of the function \fct{Reduction.Phase} is:

\begin{lstlisting}
R > out = Reduction.Phase(X = dgp$X,Y = dgp$Y,
                          family = gaussian, seed.HC = 1012)

\end{lstlisting}
In particular, this arranges the indices of the columns of \code{dgp\$X} in a hypercube of appropriate dimension, and fits a normal theory linear regression model to each set of variables indexed by the rows, columns, etc., of the hypercube. Other choices of the argument \code{family} are illustrated in \S \ref{sec:real}. The arrangement of the variable indices in the hypercube is at random. However, \code{seed.HC = 1012} allows the results of the analyses reported here to be reproduced. If the argument \code{dmHC} is left unspecified (the recommendation), as in this example, the dimension of the initial hypercube is set to be the smallest dimension such that the number of rows, columns etc.,~is no greater than 15. Thus, the present example initially has the 1000 variable indices arranged in a $10\times 10 \times 10$ cube. 

Because the comprehensive model obtained from the full data achieves better fit than an arbitrary model embedding the one to be tested, a test of adequacy of the smaller model rejects too often in hypothetical repeated application. It is therefore usually sensible to split the sample in two and use, say 70\%, for the reduction and exploratory phases, and the remaining 30\% for construction of the conditional confidence sets of models. The appropriate modification to the previous code, so that only the first 70 observations are used for the reduction phase, is:

\begin{lstlisting}
R > outSplit1 = Reduction.Phase(X = dgp$X[1:70,],
         Y = dgp$Y[1:70], family = gaussian, seed.HC = 1012)

\end{lstlisting}

If the initial sample size is rather small and the model to be fitted is non-Gaussian, the sample size available for the final phase of the procedure is likely to be too small for the distribution of the maximum likelihood estimator to be well-approximated by its asymptotic distribution. Correspondingly, the coverage probability of the confidence sets of models conditional on the reduction phase is likely to differ from the nominal value. This could be mitigated through a Bartlett correction to the likelihood ratio statistic, but this has not been implemented in the current version of the package. See Bartlett (1937), Barndorff-Nielsen and Cox (1984) and Barndorff-Nielsen and Cox (1994, p133, p152--53) for a discussion of the theory of Bartlett correction.

A strong reassurance over the security of one's conclusions is given if the set of retained variables does not alter much upon rerandomization of the arrangement of the variable indices in the (hyper)cube, and this is a suitably cautious check in practice. Indeed, if the answers so obtained differ appreciably, the suggestion is that too severe a reduction has been performed. Thus we consider also the outcome \code{outSplit2}, obtained when no argument \code{seed.HC} is provided, so that variable indices are arranged in their original order. Some variables will appear in all or almost all analyses.

\begin{lstlisting}
R > outSplit2 = Reduction.Phase(X = dgp$X[1:70,],
                         Y = dgp$Y[1:70], family = gaussian)

\end{lstlisting}

The outcomes \code{outSplit1} and \code{outSplit2} of the previous two analyses are two lists of variable indices from each successive reduction. Only the latter reductions are of ultimate interest, but the intermediate reductions should be inspected to ensure that the number of variables retained is not so large as to be detrimental to the subsequent stage of the reduction. In the present example, the final lists of variables are arrived at by an implementation of the default decision rules, to some extent guided by the analysis of Battey and Cox (2018). These are to retain variables if they are among the two most significant in at least half the analyses in which they appear in the first stage reduction, and if they are significant at the 1\% level in at least half the analysis in which they appear in subsequent reductions. The 1\% threshold is arbitrary and judgement should be exercised if the output of such an analysis is unreasonable, for instance if too many variables are retained in any stage of the reduction. This is particularly important when the initial number of variables is very large, so that variables are initially arranged in a four or five dimensional hypercube. \S \ref{sec:real} illustrates appropriate use of judgement through the optional argument \code{vector.signif} of the \fct{Reduction.Phase} function. The objective of the reduction phase is to reduce the number of candidate signal variables to ideally not more than 20, to be subjected to more detailed joint analysis.

The sorted lists of retained variables using the default decision rules and the two initial arrangements of variables indices in the cube are:

\begin{lstlisting}
R > v1=outSplit1$List.Selection$`Hypercube with dim 2`$numSelected1
R > v2=outSplit2$List.Selection$`Hypercube with dim 2`$numSelected1
\end{lstlisting}

\begin{lstlisting}
R > v1 = sort(v1)
   46 51 66 156 229 263 272 319 423 496 531 559 735 804 827 
   897 929 984 1000
R > v2 = sort(v2) 
   46 156 272 291 319 397 531 559 642 827 897 929 984
\end{lstlisting}

Of these variables, ten are in common, an appreciable overlap. The indices of the five true signal variables are contained in  \code{v1} and \code{v2} (and their intersection). These are
\begin{lstlisting}
R > dgp$TRUE.idx
   46 531 559 897 929 
\end{lstlisting}

Usage of the other functions in the package is illustrated using the output of the second analysis, i.e., the variables in \code{v2}.

An alternative to the reduction phase is to use a deliberately undertuned lasso fit. The lasso is typically fitted by the coordinate descent algorithm in general regression settings, or by the least angle regression algorithm in the linear model. Thus, the practical implementation of the lasso is essentially forward selection. By contrast, the reduction phase of Cox and Battey (2017) is a version of backward elimination. Both forward selection and backward elimination are likely to be effective in many cases, although a theoretical elucidation of the conditions on the design matrix to ensure this has not been attempted. If the objective is to obtain a superset of the comprehensive model, as here, backward elimination has advantages in simpler settings. See  \S \ref{sec:simulations} for an empirical comparison in idealized examples.

The lasso, fitted by coordinate descent as implemented in the \proglang{R} package \pkg{glmnet}, and undertuned to produce at least the same number of variables as in \code{v2}, is obtained by

\begin{lstlisting}
R > library(glmnet)
R > lasso.fit = glmnet(x = dgp$X[1:70,],y = dgp$Y[1:70])
R > n.coefs = apply(coef(lasso.fit), 2, 
                       function(x) length(which(x!=0))) 
R > idx.coefs = which(n.coefs == length(v2))
R > if(length(idx.coefs)==0){
      idx.coefs = min(which(n.coefs >= length(v2)))}
R > lasso.var = which(coef(lasso.fit)[,idx.coefs[1]]!=0) 
\end{lstlisting}
In the present example, the associated variables are

\begin{lstlisting}
R > lasso.var
   40 46  161 341 384 511 531  559  827 897  929  984
\end{lstlisting}
Seven of these are in common with \code{v1} and \code{v2}, including the five signal variables.

\subsection{Exploratory phase}

The analysis discussed by Cox and Battey (2017) is intended to be largely exploratory, and a key aspect of the procedure is that it allows informal checks, standard in much statistical work. The function \fct{Exploratory.Phase} automates some, but by no means all, of what would typically take place in an exploratory data analysis, and is provided as a rough guide. Usage of the \code{silent} argument is illustrated in \S \ref{sec:real}, in which \code{silent = FALSE} forces a certain degree of judgement to be exercised.


The following code detects potentially important squared or interaction terms among the variables whose indices are stored in \code{v2}.

\begin{lstlisting}
R > out.exp.phase = Exploratory.Phase(X = dgp$X[1:70,],
                     Y = dgp$Y[1:70], list.reduction = v2, 
                     family = gaussian, signif = 0.01)
\end{lstlisting}
Neither squared terms nor interaction terms are suggested as potentially important.

\subsection{Model selection phase}

The final stage of the procedure is to test all low-dimensional subsets of the comprehensive model for compatibility with the data. The comprehensive model is that containing all variables from the reduction phase and any squared or interaction terms suggested at the exploratory phase, of which there are none in the present example. Usage is:
\begin{lstlisting}
R > out.MS = ModelSelection.Phase(X = dgp$X[71:100,],
      Y = dgp$Y[71:100], list.reduction = v2, signif = 0.01)
\end{lstlisting}
The appropriate modification to the arguments of this function when squared or interaction terms are to be considered is illustrated in \S \ref{sec:real}.

The above finds all models of dimension 5 or smaller whose likelihood ratio test against the comprehensive is not rejected at the \code{signif = 0.01} significance level. The optional argument \code{modelSize} specifies the maximum size of the models to be searched over. The true model appears in the set of all well-fitting models identified, i.e., in the list of models displayed by the code:
\begin{lstlisting}
R > out.MS$goodModels$`Model Size 5` 
\end{lstlisting}

All models that are found to be compatible with the data should be reported. Specifically, the output of the function \fct{ModelSelection.Phase} should be used to produce (sometimes large) tables like those appearing in the supplementary file of Cox and Battey (2017), available at: \newline
\texttt{\href{https://www.pnas.org/content/pnas/suppl/2017/07/20/1703764114.DCSupplemental}{https://www.pnas.org/content/pnas/suppl/2017/07/20/1703764114.DCSupplemental}}

Provided that the sample is split as in the example above, such tables constitute a conditional confidence set of models. Conditional on the reduction phase, these have, in principle, exact nominal coverage in the normal theory linear regression model and asymptotically nominal coverage in more general regression models fitted by maximum likelihood.

\section{Illustration of performance in some idealized settings} \label{sec:simulations}

The present section explores empirical sensitivities of the procedure to modifications to the data generating mechanism. Several aspects are of interest: sensitivity of the reduction phase as described by Cox and Battey (2017) (a version of backward elimination) and of the undertuned lasso (a version of forward selection) in terms of retaining the true model in its entirety; efficacy of the model confidence sets in terms of their coverage probabilities and size. Full sample and split sample properties of both approaches are considered.

It is an open problem to elucidate the conditions on the design matrix and signal strength in order for the procedure based on traversal of successively lower dimensional hypercubes to retain a reasonably sized superset of the true set of signal variables with quantifiable high probability. Some related discussion for the undertuned lasso is given by B{\"u}hlmann and van de Geer (2011, chapter 7) and Belloni and Chernozhukov (2013). 

In the tables below, $\mathcal{S}$ is the true set of signal variables, $\widehat{\mathcal{S}}$ is the set of variables surviving the reduction phase, $\mathcal{M}$ is the set of low-dimensional models whose likelihood ratio test against the comprehensive model is not rejected at the 1\% level. In all the simulation experiments considered, the first stage of the reduction phase arranges the 1000 variables in a $10\times10\times 10$ cube and retains variables if they are among the two most significant in at least two of the three analyses in which they appear. The second stage reduction is tuned so that approximately 10-20 variables are retained through the reduction phase, however the associated threshold for the significance tests is fixed across Monte Carlo replications so that the number of retained variables is random. Results for the normal theory linear model with a sample of size $n=100$ are reported in Table 1, where `CB' is the procedure of Cox and Battey(2017)  implemented using `HCmodelSets'. The threshold of the second-stage significance test is 0.1\%.

\bigskip

{\centering

\begin{table}[h]
\resizebox{\textwidth}{!}{\begin{tabular}{cccc|cccc|cc|cc}
& & & & \multicolumn{4}{c|}{pr($\mathcal{S}\subseteq \widehat{\mathcal{S}}$)} &\multicolumn{2}{c|}{pr($\mathcal{S}\in \mathcal{M}$)} & \multicolumn{2}{c}{$\mathbb{E}|\mathcal{M}\backslash \mathcal{S}|$}  \\

\multirow{2}{*}{$v_{S0}$} & \multirow{2}{*}{$v_{C0}$} & \multirow{2}{*}{$\rho$} & \underline{signal} & undertuned & undertuned & CB & CB & CB & CB & CB & CB \\
& & & noise & lasso (full) & lasso (split) & (full) & (split)  & (full)  & (split) & (full) & (split)  \\
\hline
1 & 1 & 0.9 & 1   & 1.00 (0.04) & 0.99 (0.08) & 1.00 (0.00) & 1.00 (0.00) & 0.57 (0.49) & 0.98 (0.14) & 6.16 (7.25)    & 16.3 (28.6)   \\

1 & 1 & 0.9 & 0.6 & 0.94 (0.24) & 0.84 (0.37) & 0.98 (0.15) & 0.83 (0.37) & 0.41 (0.49) & 0.83 (0.38) & 4.93 (4.82)    & 16.1 (31.5)   \\

1 & 1 & 0.5 & 1   & 0.92 (0.27) & 0.87 (0.34) & 1.00 (0.00) & 1.00 (0.00) & 0.56 (0.50) & 0.98 (0.13) & 4.63 (5.89)    & 8.73 (18.7)    \\

1 & 1 & 0.5 & 0.6 & 0.85 (0.36) & 0.75 (0.43) & 0.99 (0.11) & 0.85 (0.35) & 0.39 (0.49) & 0.84 (0.36) & 2.29 (2.73)    & 8.68 (19.2)    \\ \hline

1 & 3 & 0.9 & 1   & 1.00 (0.04) & 0.99 (0.08) & 1.00 (0.04) & 0.99 (0.08) & 0.62 (0.49) & 0.96 (0.19) & 24.6 (24.8)  & 77.2 (126)  \\

1 & 3 & 0.9 & 0.6 & 0.92 (0.27) & 0.79 (0.41) & 0.97 (0.16) & 0.81 (0.39) & 0.48 (0.50) & 0.79 (0.41) & 22.7 (18.4)  & 44.6 (78.1)   \\

1 & 3 & 0.5 & 1   & 0.97 (0.16) & 0.92 (0.27) & 1.00 (0.00) & 1.00 (0.00) & 0.59 (0.49) & 0.98 (0.13) & 10.9 (14.0)  & 14.3 (36.1)   \\

1 & 3 & 0.5 & 0.6 & 0.89 (0.31) & 0.82 (0.39) & 0.97 (0.17) & 0.87 (0.34) & 0.36 (0.48) & 0.85 (0.35) & 3.66 (5.01)    & 10.3 (25.0)   \\ \hline

5 & 1 & 0.9 & 1   & 0.98 (0.15) & 0.95 (0.21) & 1.00 (0.00) & 1.00 (0.00) & 0.94 (0.23) & 0.98 (0.13) & 7.15 (7.19)    & 80.4 (85.7)   \\

5 & 1 & 0.9 & 0.6 & 0.79 (0.40) & 0.57 (0.50) & 1.00 (0.04) & 0.98 (0.15) & 0.89 (0.31) & 0.96 (0.19) & 40.9 (35.7)  & 146 (149) \\

5 & 1 & 0.5 & 1   & 1.00 (0.04) & 0.98 (0.15) & 1.00 (0.00) & 1.00 (0.00) & 0.96 (0.20) & 0.99 (0.11) & 0.01 (0.10)    & 8.64 (13.0)    \\

5 & 1 & 0.5 & 0.6 & 0.99 (0.10) & 0.96 (0.21) & 1.00 (0.00) & 0.99 (0.10) & 0.88 (0.32) & 0.98 (0.15) & 1.18 (2.04)    & 51.6 (64.2)   \\ \hline

5 & 3 & 0.9 & 1   & 0.99 (0.11) & 0.95 (0.22) & 1.00 (0.00) & 1.00 (0.00) & 0.94 (0.24) & 0.98 (0.15) & 16.7 (18.4)  & 212 (202) \\

5 & 3 & 0.9 & 0.6 & 0.77 (0.42) & 0.51 (0.50) & 1.00 (0.06) & 0.96 (0.20) & 0.86 (0.35) & 0.94 (0.24) & 101 (88.2) & 418 (351) \\

5 & 3 & 0.5 & 1   & 1.00 (0.00) & 1.00 (0.00) & 1.00 (0.00) & 1.00 (0.00) & 0.96 (0.20) & 0.98 (0.14) & 0.01 (0.10)    & 20.0 (35.0)   \\

5 & 3 & 0.5 & 0.6 & 1.00 (0.06) & 0.98 (0.13) & 1.00 (0.00) & 0.99 (0.12) & 0.89 (0.32) & 0.96 (0.20) & 2.85 (4.03)    & 123 (137) \\ \hline
\end{tabular}}
\medskip
\caption{Monte Carlo estimates and their estimated standard errors (in parentheses) from 500 Monte Carlo draws from the normal theory linear model with parameter combinations as displayed. In the split sample case, 70 observations are used for reduction and 30 for construction of the confidence sets of models.}
\end{table}

}

The same experiment is performed on survival time data, generated according to a proportional hazards model with Weibull baseline hazard as described in \S \ref{sec:dgp}. The survival times are censored, with the censoring times generated from an exponential distribution of rate $0.1$. In particular, the code fragment of \S \ref{sec:dgp} is modified so that in each Monte Carlo replication, data are generated as: 

\begin{lstlisting}

R > dgp = DGP(s=5, a=3, sigStrength=1, rho=0.9, n=100, 
    intercept=5, var=1, d=1000, type.response = "S", scale=1, 
    shape=1, rate=0.1, DGP.seed=2018)        
\end{lstlisting}

In the notation of \S \ref{sec:dgp}, the parameters of the Weibull distribution are set as $\tau$\code{ = scale = 1}, and $\kappa$\code{ = shape = 1}. Knowledge of the baseline hazard is ignored and the data are fitted by partial likelihood as implemented in the \fct{coxph} function of the `survival' package, available at \url{https://cran.r-project.org/web/packages/survival/index.html}. Summary statistics over 500 Monte Carlo replications are reported in Table 2. The threshold of the second-stage significance test is 0.25\%.

\bigskip

{\centering

\begin{table}[h]
\resizebox{\textwidth}{!}{\begin{tabular}{cccc|cccc|cc|cc}
& & & & \multicolumn{4}{c|}{pr($\mathcal{S}\subseteq \widehat{\mathcal{S}}$)} &\multicolumn{2}{c|}{pr($\mathcal{S}\in \mathcal{M}$)} & \multicolumn{2}{c}{$\mathbb{E}|\mathcal{M}\backslash \mathcal{S}|$}  \\

\multirow{2}{*}{$v_{S0}$} & \multirow{2}{*}{$v_{C0}$} & \multirow{2}{*}{$\rho$} & \underline{signal} & undertuned & undertuned & CB & CB & CB & CB & CB & CB \\
& & & noise & lasso (full) & lasso (split) & (full) & (split)  & (full)  & (split) & (full) & (split)  \\
\hline
1 & 1 & 0.9 & 1   & 1.00 (0.00) & 1.00 (0.00) & 1.00 (0.00) & 1.00 (0.00) & 0.03 (0.17) & 0.95 (0.23) & 54.1 (92.2)   & 1273 (1490) \\
1 & 1 & 0.9 & 0.6 & 0.99 (0.12) & 0.94 (0.24) & 1.00 (0.04) & 0.97 (0.17) & 0.00 (0.04) & 0.89 (0.31) & 15.6 (57.3)   & 1863 (2264) \\
1 & 1 & 0.5 & 1   & 1.00 (0.00) & 1.00 (0.00) & 1.00 (0.00) & 1.00 (0.00) & 0.04 (0.21) & 0.95 (0.21) & 57.4 (97.4)   & 962 (1085)   \\
1 & 1 & 0.5 & 0.6 & 1.00 (0.00) & 0.98 (0.13) & 0.99 (0.08) & 0.96 (0.20) & 0.00 (0.00) & 0.90 (0.31) & 13.0 (31.6)   & 1734 (2374) \\ \hline
1 & 3 & 0.9 & 1   & 1.00 (0.00) & 0.99 (0.09) & 1.00 (0.00) & 1.00 (0.04) & 0.07 (0.25) & 0.95 (0.22) & 102 (209) & 2468 (2738) \\
1 & 3 & 0.9 & 0.6 & 0.97 (0.18) & 0.90 (0.30) & 0.98 (0.13) & 0.95 (0.21) & 0.01 (0.09) & 0.91 (0.29) & 45.0 (98.5)   & 3182 (3700) \\
1 & 3 & 0.5 & 1   & 1.00 (0.00) & 1.00 (0.00) & 1.00 (0.00) & 1.00 (0.00) & 0.07 (0.25) & 0.95 (0.22) & 105 (158) & 1094 (1090) \\
1 & 3 & 0.5 & 0.6 & 1.00 (0.00) & 1.00 (0.06) & 1.00 (0.00) & 0.97 (0.17) & 0.00 (0.04) & 0.91 (0.28) & 18.6 (51.1)   & 1955 (2859) \\ \hline
5 & 1 & 0.9 & 1   & 0.98 (0.15) & 0.90 (0.30) & 1.00 (0.00) & 1.00 (0.04) & 0.78 (0.41) & 0.91 (0.29) & 30.9 (46.5)   & 916 (1165)   \\
5 & 1 & 0.9 & 0.6 & 0.79 (0.41) & 0.52 (0.50) & 1.00 (0.00) & 0.99 (0.08) & 0.59 (0.49) & 0.94 (0.24) & 136 (180) & 2216 (2390) \\
5 & 1 & 0.5 & 1   & 1.00 (0.00) & 1.00 (0.00) & 1.00 (0.00) & 1.00 (0.00) & 0.80 (0.40) & 0.91 (0.28) & 0.00 (0.09)     & 59.0 (118)      \\
5 & 1 & 0.5 & 0.6 & 1.00 (0.00) & 0.99 (0.12) & 1.00 (0.00) & 1.00 (0.04) & 0.54 (0.50) & 0.90 (0.31) & 1.46 (4.22)     & 382 (572)     \\ \hline
5 & 3 & 0.9 & 1   & 0.98 (0.13) & 0.86 (0.35) & 1.00 (0.00) & 1.00 (0.04) & 0.80 (0.40) & 0.86 (0.35) & 46.4 (66.2)   & 1383 (1682) \\
5 & 3 & 0.9 & 0.6 & 0.71 (0.45) & 0.48 (0.50) & 1.00 (0.00) & 0.99 (0.11) & 0.63 (0.48) & 0.90 (0.30) & 242 (310) & 2846 (2603) \\
5 & 3 & 0.5 & 1   & 1.00 (0.00) & 1.00 (0.00) & 1.00 (0.00) & 1.00 (0.00) & 0.83 (0.38) & 0.87 (0.34) & 0.09 (1.03)     & 73.4 (175)      \\
5 & 3 & 0.5 & 0.6 & 1.00 (0.00) & 0.99 (0.11) & 1.00 (0.00) & 1.00 (0.04) & 0.59 (0.49) & 0.90 (0.30) & 2.35 (5.25)     & 575 (925)     \\ \hline
\end{tabular}}
\medskip
\caption{Monte Carlo estimates and their estimated standard errors (in parentheses) from 500 Monte Carlo draws from the Weibull proportional hazards model with parameter combinations as displayed.}
\end{table}

}

The results are qualitatively similar to those for the normal theory linear model. The main difference is that the coverage probability of the confidence sets of models, conditional on all variables being retained through the first stage reduction, is lower than the 0.99 nominal level. The reason is that the distribution theory underpinning the associated likelihood ratio tests is, in principle, exact for the normal theory linear model and is at best asymptotically valid for most other types of regression model. This could be mitigated through a Bartlett correction to the likelihood ratio statistic, but this has not been implemented in the current version of the package. See Bartlett (1937), Barndorff-Nielsen and Cox (1984) and Barndorff-Nielsen and Cox (1994, p133, p152--53) for a discussion of the theory of Bartlett correction. The results of Table 2 are for $n=150$ with 100 observations used for reduction and 50 for construction of confidence sets of models in the split sample case.

As mentioned previously, adjustments to the likelihood ratio statistic to improve the $\chi^2$ approximation to its distribution are possible, but these have not been implemented in `HCmodelSets'.

\section{Real example} \label{sec:real}

We now illustrate the use of `HCmodelSets' to construct conditional confidence sets of models for the survival times of lymphoma patients. The data, which can be downloaded from
\href{https://www.jstatsoft.org/article/view/v039i05}{https://www.jstatsoft.org/article/view/v039i05}, are from the study of Alizadeh \emph{et al.} (2000) and also used by Simon \emph{et al.} (2011). There are measurements on $d=7399$ genetic variants for $n=240$ patients. The indices of these variables are arranged in a 4 dimensional hypercube, which is the default starting dimension. As before, the data are divided into those to be used in the reduction and exploratory phases and those to be used in the model selection phases.
\begin{lstlisting}
R > attach("LymphomaData.rda")
R > x = t(patient.data$x)
R > y = patient.data$time
R > status = patient.data$status
# Data Splitting
R > X.in = x[1:168,]
R > Y.in = y[1:168]
R > status.in = status[1:168]
R > Y = cbind(Y.in,status.in)
R > X.out = x[169:240,]
R > Y.out = y[169:240]
R > status.out = status[169:240] 
\end{lstlisting}

The first stage decision rule is to retain all variables that are among the two most significant in at least two of the three analyses in which they appear. The decision rules for the remaining reduction stages are specified by the argument \code{vector.signif} in the \fct{Reduction.Phase} function:

\begin{lstlisting}
R > library(HCmodelSets)
R > out.1 = Reduction.Phase(X = X.in,Y = Y,Cox.Hazard = TRUE,
              vector.signif = c(2,0.0025,0.001), seed.HC = 2)
\end{lstlisting}

The choice 
\code{vector.signif = c(2,0.0025,0.001)} means that the second stage decision rule retains variables if they are significant at the 0.25\% level in at least two of the three analyses in which they appear and the third stage decision rule retains variables if they are significant at the 0.1\% level in at least one of the two analyses in which they appear. This choice was determined by checking that the numbers of variables retained through each stage of the reduction is sensible, that the number of variables ultimately retained is within the target range, and that the outcome is not too sensitive to changes to the original arrangement of the variable indices in the hypercube. The set of variables ultimately retained is
\begin{lstlisting}
R > v1 = out.1$List.Selection$`Hypercube with dim 2`$numSelected1
R > sort(v1)
   1188 1660 1825 2437 2879 2902 3172 3177 3800 3814 3822 3824 
   5027 6134 6706 6896 7357 
\end{lstlisting}
Rerandomizing the variable indices in the hypercube produces the set of variables
\begin{lstlisting}
R > out.2 = Reduction.Phase(X = X.in,Y = Y,Cox.Hazard = TRUE,
           vector.signif = c(2,0.0025,0.001), seed.HC = 11)
R > v2 = out.2$List.Selection$`Hypercube with dim 2`$numSelected1           
\end{lstlisting}
\begin{lstlisting}
R> sort(v2)
  1188 1675 1714 1825 1984 2437 2900 3172 3800 3811 3818 
  3819 3822 3833  4126  5027  6134  6706  7069  7357 
\end{lstlisting}
Ten of the variables in the original list of 17 are also in the second list. The lasso, undertuned to select at least the same number of variables as in \code{v1} produces an overlap of 8 variables with \code{v1}, namely,

\begin{lstlisting}
R > library(glmnet)
R > lasso.fit = glmnet(X.in, Surv(Y.in,status.in), 
	        family = "cox", alpha = 1)
R > n.coefs = apply(coef(lasso.fit), 2, 
            function(x) length(which(x!=0)))
R > idx.coefs = which(n.coefs == length(v1))
R > if(length(idx.coefs)==0){
	        idx.coefs = min(which(n.coefs >= v1))}
R > lasso.var = which(coef(lasso.fit)[,idx.coefs[1]]!=0)
\end{lstlisting}
\begin{lstlisting}
R > lasso.var
  394 1072 1188 1456 1662 1681 1825 2902 3172 3180 3801
  3822 4882 5027 6134 6896 7357 
\end{lstlisting}
The variable 3801, found by the lasso, has empirical correlation greater than 0.9 with variable 3800 in \code{v1} and \code{v2}.

The exploratory phase now uses significance tests as an informal guide to suggesting potential squared or interaction terms. For each of the variables in \code{v1}, a regression is fitted by partial likelihood with its squared term added. Extreme $t$ statistics on squared terms suggest a potentially important effect. The the linear by linear interactions of pairs of variables are checked in a similar way, with \code{silent = FALSE} in \fct{Exploratory.Phase} producing plots of the response variable as a function of pairs of variables for any interaction suggested as potentially important. Example usage is

\begin{lstlisting}
R > out.exp.phase =  Exploratory.Phase(X=X.in,Y=Y,
	 		list.reduction = v1, silent = FALSE, 
	 		Cox.Hazard = TRUE, signif=0.01)                                           
\end{lstlisting}
which produces a sequence of plots and questions of the form\newline 
\begin{center} 
\vspace{-0.5cm}
\code{Discard interaction term? [Y/N]}.\newline
\vspace{-0.5cm}
\end{center}
For illustrative purposes, we answer \code{N} (no) to the questions for which the plots are displayed in Figure 1, although the suggestion from an interaction plot ought to be much stronger to justify an interaction's inclusion. See Cox and Battey (2017) for an example.

\begin{figure*}[]
\centering
\includegraphics[height=5cm,width=6.5cm]{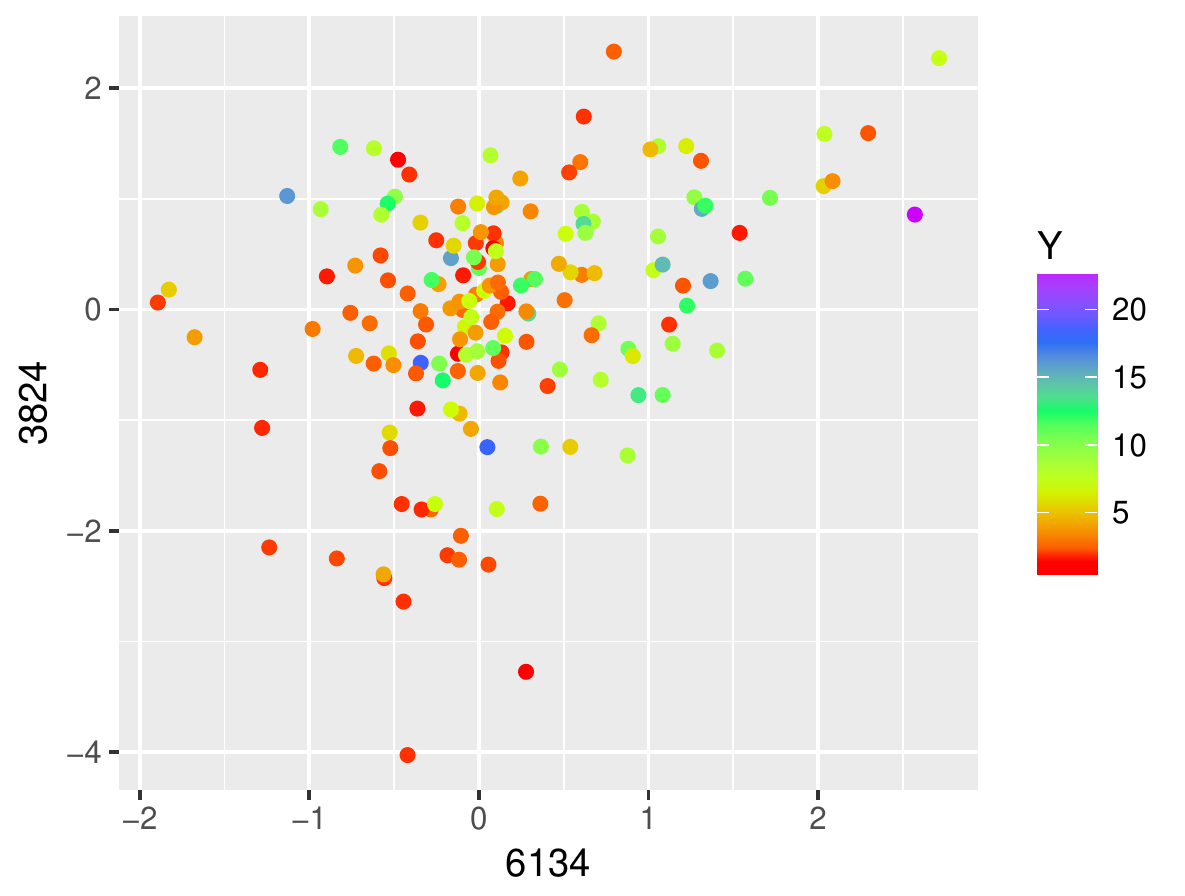}
\includegraphics[height=5cm,width=6.5cm]{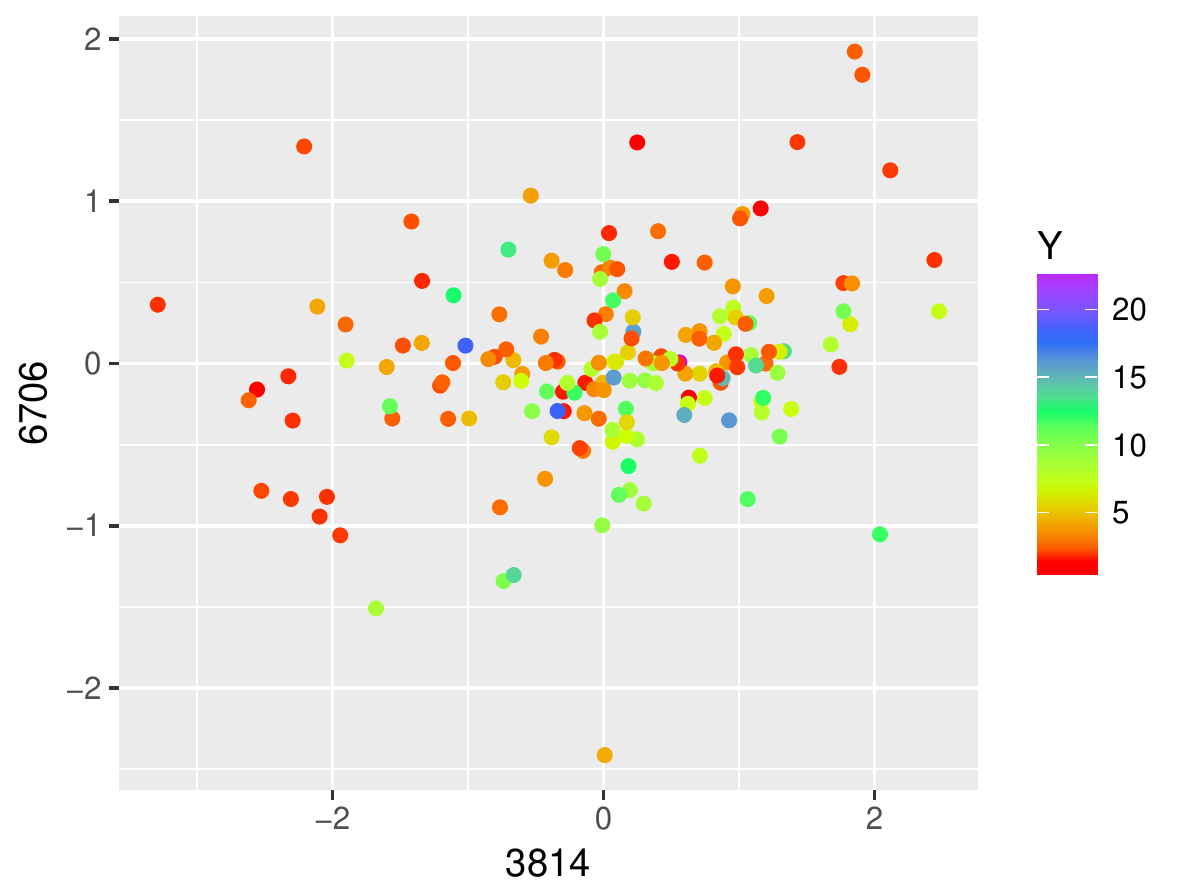}
\caption{Interaction plots of variables (6134, 3824) and (3814, 6706)}
\label{iteractions}
\end{figure*}

Thus we have 20 variables in all, the seventeen variables contained in \code{v1}, one squared term contained in \code{out.Exploratory.Phase\$mat.select.SQ} and two interaction terms given by the rows of \code{out.Exploratory.Phase\$mat.select.INTER}. The analysis proceeds as follows:
\begin{lstlisting}
R > sq.terms = out.exp.phase$mat.select.SQ
R > in.terms = out.exp.phase$mat.select.INTER
R > out.MS = ModelSelection.Phase(X = X.out,
   Y = cbind(Y.out,status.out), 
   list.reduction = v1,Cox.Hazard = TRUE, sq.terms = sq.terms, 
   in.terms = in.terms,signif = 0.05, modelSize = 7)
\end{lstlisting}
Sets of well-fitting models of different sizes, up to \code{modelSize = 7}, are contained in the list \code{out.MS\$goodModels}. For instance, 

\begin{lstlisting}
out.MS$goodModels$`Model Size 2`
\end{lstlisting}
produces a list of well-fitting models of size 2. If there are models for which an interaction term is present without the corresponding main effects, the main effects are added. Thus, there are 23 models of size 2, statistically indistinguishable from the comprehensive model at the 5\% significance level.

Of all the well-fitting models identified 72\% involve the variable 3824 and 70\% involve the variable 6134. A very small proportion of models contain neither 3814 nor 6134. Indeed variables 3814 and 6134 occur frequently, but rarely together. Table 3 reports the proportion of models containing variable A, given that they do not contain variable B, say. While one should be cautious over overinterpretting the output, these give an indication of which variables might be substitutes for one another. The variables have been ordered, from left to right and from top to bottom, in order of their frequency of appearance in the sets of models. For typographical reasons their indices have been recoded as: 1=1188; 2=1660; 3=1825; 4=2437; 5=2879; 6=2902; 7=3172; 8=3177; 9=3800; 10=3814; 11=3822; 12=3824; 13=5027; 14=6134; 15=6706; 16=6896; 17=7357; 18=squared term on 3814; 19=interaction between 6134 and 3824; 20=interaction between 3814 and 6706.

For an example of other summary tables of potential interest, see the supplementary file of Cox and Battey (2017).

\bigskip

{\centering

\begin{table}
\resizebox{\textwidth}{!}{\begin{tabular}{c|cccccccccccccccccccc}
A	 & \multicolumn{20}{c}{B} \\
   & 12   & 14   & 19   & 10   & 4    & 15   & 20   & 3    & 7    & 9    & 18   & 1    & 11   & 17   & 13   & 5    & 6    & 16   & 2    & 8    \\
   \hline
12 &      & 0.26 & 0.24 & 0.87 & 0.81 & 0.84 & 0.76 & 0.75 & 0.77 & 0.74 & 0.72 & 0.73 & 0.73 & 0.72 & 0.72 & 0.71 & 0.71 & 0.71 & 0.71 & 0.70 \\
14 & 0.22 &      & 0.20 & 0.85 & 0.80 & 0.83 & 0.75 & 0.74 & 0.75 & 0.73 & 0.71 & 0.72 & 0.71 & 0.70 & 0.71 & 0.70 & 0.69 & 0.69 & 0.69 & 0.69 \\
19 & 0.00 & 0.00 &      & 0.82 & 0.75 & 0.79 & 0.69 & 0.67 & 0.70 & 0.65 & 0.63 & 0.65 & 0.63 & 0.62 & 0.63 & 0.61 & 0.61 & 0.61 & 0.61 & 0.60 \\
10 & 0.75 & 0.73 & 0.73 &      & 0.47 & 0.35 & 0.37 & 0.45 & 0.46 & 0.44 & 0.47 & 0.47 & 0.48 & 0.47 & 0.47 & 0.48 & 0.48 & 0.48 & 0.47 & 0.48 \\
4  & 0.62 & 0.63 & 0.63 & 0.45 &      & 0.41 & 0.48 & 0.48 & 0.40 & 0.46 & 0.47 & 0.43 & 0.43 & 0.45 & 0.44 & 0.43 & 0.44 & 0.44 & 0.45 & 0.44 \\
15 & 0.68 & 0.68 & 0.68 & 0.31 & 0.40 &      & 0.31 & 0.42 & 0.43 & 0.43 & 0.42 & 0.43 & 0.43 & 0.44 & 0.44 & 0.44 & 0.44 & 0.45 & 0.44 & 0.44 \\
20 & 0.48 & 0.47 & 0.48 & 0.28 & 0.43 & 0.26 &      & 0.40 & 0.40 & 0.40 & 0.40 & 0.39 & 0.39 & 0.39 & 0.40 & 0.39 & 0.39 & 0.39 & 0.39 & 0.39 \\
3  & 0.44 & 0.44 & 0.44 & 0.34 & 0.41 & 0.36 & 0.38 &      & 0.38 & 0.39 & 0.37 & 0.37 & 0.37 & 0.37 & 0.36 & 0.37 & 0.37 & 0.37 & 0.38 & 0.37 \\
7  & 0.48 & 0.48 & 0.48 & 0.37 & 0.32 & 0.37 & 0.38 & 0.38 &      & 0.39 & 0.37 & 0.37 & 0.38 & 0.38 & 0.38 & 0.38 & 0.37 & 0.37 & 0.36 & 0.37 \\
9  & 0.38 & 0.38 & 0.38 & 0.30 & 0.35 & 0.34 & 0.35 & 0.35 & 0.35 &      & 0.35 & 0.34 & 0.32 & 0.34 & 0.34 & 0.34 & 0.33 & 0.34 & 0.34 & 0.34 \\
18 & 0.33 & 0.34 & 0.32 & 0.33 & 0.37 & 0.32 & 0.33 & 0.33 & 0.32 & 0.34 &      & 0.34 & 0.33 & 0.33 & 0.33 & 0.33 & 0.33 & 0.33 & 0.33 & 0.33 \\
1  & 0.35 & 0.35 & 0.35 & 0.32 & 0.31 & 0.32 & 0.32 & 0.32 & 0.32 & 0.32 & 0.33 &      & 0.32 & 0.32 & 0.33 & 0.32 & 0.32 & 0.32 & 0.32 & 0.32 \\
11 & 0.32 & 0.31 & 0.30 & 0.32 & 0.28 & 0.29 & 0.30 & 0.30 & 0.31 & 0.29 & 0.30 & 0.31 &      & 0.30 & 0.30 & 0.30 & 0.31 & 0.30 & 0.30 & 0.30 \\
17 & 0.29 & 0.29 & 0.28 & 0.30 & 0.30 & 0.30 & 0.29 & 0.30 & 0.30 & 0.30 & 0.29 & 0.30 & 0.30 &      & 0.29 & 0.30 & 0.30 & 0.30 & 0.30 & 0.30 \\
13 & 0.30 & 0.30 & 0.28 & 0.30 & 0.29 & 0.30 & 0.31 & 0.28 & 0.30 & 0.30 & 0.30 & 0.31 & 0.30 & 0.29 &      & 0.29 & 0.30 & 0.29 & 0.30 & 0.30 \\
5  & 0.25 & 0.25 & 0.24 & 0.28 & 0.26 & 0.28 & 0.27 & 0.27 & 0.28 & 0.28 & 0.27 & 0.27 & 0.28 & 0.28 & 0.27 &      & 0.28 & 0.27 & 0.27 & 0.27 \\
6  & 0.24 & 0.23 & 0.22 & 0.28 & 0.26 & 0.27 & 0.26 & 0.26 & 0.26 & 0.26 & 0.27 & 0.27 & 0.27 & 0.27 & 0.27 & 0.27 &      & 0.27 & 0.27 & 0.27 \\
16 & 0.23 & 0.23 & 0.22 & 0.27 & 0.25 & 0.28 & 0.26 & 0.26 & 0.25 & 0.26 & 0.26 & 0.26 & 0.26 & 0.26 & 0.26 & 0.26 & 0.26 &      & 0.26 & 0.26 \\
2  & 0.23 & 0.23 & 0.21 & 0.27 & 0.27 & 0.27 & 0.26 & 0.27 & 0.25 & 0.26 & 0.26 & 0.26 & 0.26 & 0.26 & 0.26 & 0.26 & 0.26 & 0.26 &      & 0.26 \\
8  & 0.20 & 0.20 & 0.18 & 0.26 & 0.24 & 0.26 & 0.24 & 0.24 & 0.24 & 0.24 & 0.24 & 0.24 & 0.24 & 0.24 & 0.25 & 0.24 & 0.24 & 0.25 & 0.25 &     \\
\hline
\end{tabular}}
\medskip
\caption{The proportion of the models in the confidence set not containing variable $B$ that contain variable $A$, i.e. $|\mathcal{M}(A\cap \neg B)|/|\mathcal{M}(\neg B)|$, where $\mathcal{M}(\neg B)$ is the set of models in the confidence set that do not contain variable $B$.}
\end{table}

}

\section{Summary} \label{sec:summary}

In the context of regression with a large number of potential explanatory variables Cox and Battey (2017) emphasize that if there are several statistically indistinguishable explanations of the data, one should aim to specify as many as is feasible, a view that is in contraposition to that implicit in the use of the lasso and similar methods. The approach of Cox and Battey (2017) entails reducing the set of variables to those that potentially have an individual effect on the response, followed by more detailed joint exploration, requiring judgement at various stages. We have discussed the \proglang{R} implementation of these new ideas in `HCmodelSets'. Matlab code is also available at \url{http://wwwf.imperial.ac.uk/~hbattey/softwareCube.html}.

\subsubsection*{Acknowledgments}

We thank D.~R.~Cox for valuable comments and M.~Avella Medina for a helpful reference. The work of H.~H.~Hoeltgebaum was fully supported by the National Council for Research and Development, CNPq, Ministry of Science and Technology, Brazil. The development of `HCmodelSets' and the work of H.~S.~Battey was supported by a UK Engineering and Physical Sciences Research Fellowship (grant number EP/P002757/1).

\bigskip
\bigskip
\bigskip
\bigskip

{\centering

REFERENCES

}

\bigskip
\bigskip

Alizadeh, et al (2000). Distinct types of diffuse large B-cell lymphoma identified by gene expression profiling. \emph{Nature}, 403. \\

Barndorff-Nielsen, O.~and Cox~D.~R.~(1984). Bartlett adjustments to the likelihood ratio statistic and the distribution of the maximum likelihood estimator. \emph{J.~R.~Statist.~Soc. B}, 483--495. \\

Barndorff-Nielsen, O.~and Cox, D.~R.~(1994). \emph{Inference and Asymptotics}. Chapman \& Hall. \\

Bartlett, M.~S.~(1937). Properties of Sufficiency and Statistical Tests. \emph{Proc. R. Soc. Lond. A}, 268--282. \\

Battey, H.~S.~and Cox, D.~R.~(2018). Large numbers of explanatory variables: a probabilistic assessment. \emph{Proc.~R.~Soc.~A}, 474. \\

Belloni, A.~and Chernozhukov, V.~(2013). Least squares after model selection in high-dimensional sparse models. \emph{Bernoulli}, 521--547. \\

B{\"u}hlmann, P.~and van de Geer, S.~(2011). \emph{Statistics for high-dimensional data}. Springer, Heidelberg. \\

Cox, D.~R~(1958). The regression analysis of binary sequences (with discussion). \emph{J.~R.~Statist.~Soc. B}, 20, 215--242. \\

Cox, D.~R.~(1972). Regression models and life-tables (with discussion). \emph{J. R. Statist. Soc. B}, 34, 187--220. \\

Cox, D.~R.~(1975a). A note on data-splitting for the evaluation of significance levels. \emph{Biometrika}, 62, 441--444. \\

Cox, D.~R.~(1975b). Partial likelihood. \emph{Biometrika}, 62, 269--276. \\

Cox, D.~R.~and Battey, H.~S.~(2017). Large numbers of explanatory variables, a semi-descriptive analysis. \emph{Proc.~Nat.~Acad.~Sci.}, 114, 8592--8595. \\

Hoeltgebaum, H.~H.~(2018). HCmodelSets: regression with a large number of potential explanatory variables. R package version 1.0.2. \\

Simon, N., Friedman, J., Hastie, T., Tibshirani, R.~(2011). Regularization paths for Cox’s pro- portional hazards model via coordinate descent. \emph{J. Statist. Software}, 39. \\

Tibshirani, R.~(1996). Regression shrinkage and selection via the lasso. \emph{J. R. Statist. Soc. B}, 58, 267--288. \\

Yates, F.~(1936). A new method of arranging variety trials involving a large number of varieties. \emph{J. Agric. Sci.}, 26, 424–455.

\end{document}